\documentclass{raa}            

\usepackage{graphicx,times}             
\usepackage{epstopdf}
\usepackage{natbib}
\usepackage{amssymb}
\usepackage{amsmath}

\begin{document}
\title{Laboratory study of the formation of fullerene (from smaller to larger, C$_{44}$ to C$_{70}$)/anthracene cluster cations in the gas phase
}

   \volnopage{Vol.0 (20xx) No.0, 000--000}      
   \setcounter{page}{1}          

   \author{Deping Zhang
      \inst{1,2}
   \and Yuanyuan Yang
      \inst{1,2,3}
   \and Xiaoyi Hu
      \inst{1,2,3}
   \and Junfeng Zhen
      \inst{1,2}
   }

   \institute{CAS Key Laboratory for Research in Galaxies and Cosmology, Department of Astronomy, University of Science and Technology of China, Hefei 230026, China; {\it dpzhang@ustc.edu.cn; jfzhen@ustc.edu.cn}\\
        \and
             School of Astronomy and Space Science, University of Science and Technology of China, Hefei 230026, China\\
        \and
            CAS Center for Excellence in Quantum Information and Quantum Physics, Hefei National Laboratory
            for Physical Sciences at the Microscale, and Department of Chemical Physics, University of Science and
            Technology of China, Hefei 230026, China\\
\vs\no
   {\small Received~~20xx month day; accepted~~20xx~~month day}}

\abstract{
The formation and evolution mechanism of fullerenes in the planetary nebula or in the interstellar medium are still not understood. Here we present the study on the cluster formation and the relative reactivity of fullerene cations (from smaller to larger, C$_{44}$ to C$_{70}$) with anthracene molecule (C$_{14}$H$_{10}$). The experiment is performed in the apparatus that combines a quadrupole ion trap with a time-of-flight mass spectrometer. By using a 355 nm laser beam to irradiate the trapped fullerenes cations (C$_{60}$$^+$ or C$_{70}$$^+$), smaller fullerene cations C$_{(60-2n)}$$^+$, n=1-8 or C$_{(70-2m)}$$^+$, m=1-11 are generated, respectively. Then reacting with anthracene molecules, series of fullerene/anthracene cluster cations are newly formed (e.g., (C$_{14}$H$_{10}$)C$_{(60-2n)}$$^+$, n=1-8 and (C$_{14}$H$_{10}$)C$_{(70-2m)}$$^+$, m=1-11), and slight difference of the reactivity within the smaller fullerene cations are observed. Nevertheless, smaller fullerenes show obviously higher reactivity when comparing to fullerene C$_{60}$$^+$ and C$_{70}$$^+$. \\
A successive loss of C$_2$ fragments mechanism is suggested to account for the formation of smaller fullerene cations, which then undergo addition reaction with anthracene molecules to form the fullerene-anthracene cluster cations. It is found that the higher laser energy and longer irradiation time are key factors that affect the formation of smaller fullerene cations. This may indicate that in the strong radiation field environment (such as photon-dominated regions) in space, fullerenes are expected to follow the top-down evolution route, and then form small grain dust (e.g., clusters) through collision reaction with co-existing molecules, here, smaller PAHs. 
}

   \authorrunning{Zhang et al. 2020}            
   \titlerunning{Formation of fullerene/anthracene cluster cations }  

  \maketitle
%

\section{Introduction}           
\label{sect:intro}
Polycyclic aromatic hydrocarbons (PAHs) are well recognized as an essential component of the interstellar medium (ISM) and may account for $ \leq $15\% of the interstellar carbons. They are observed via the infrared (IR) emission bands at 3.3, 6.2, 7.7, 8.6, 11.3 and 12.7~$\mu$m widespread throughout the Universe \citep{allamandola_interstellar_1989,genzel_what_1998,sellgren_near-infrared_1984,tielens_interstellar_2008,tielens_molecular_2013-1,li_spitzers_2020-1}. A huge effort was undertaken in the past decades to identify the carriers of those IR emission features, however, no specific PAH responsible for the IR emission features has been identified.
By contrast, another important type of carbonaceous species, fullerenes (C$_{60}$ and C$_{70}$) have been unambiguously observed in planetary nebula Tc1 via their IR emission spectra \citep{cami_detection_2010-1}. 
After that, C$_{60}$ has also been detected in reflection nebulae\citep{sellgren_c_2010,peeters_1520_2012,boersma_spatial_2012}, protoplanetary nebulae\citep{zhang_detection_2011-1}, R Coronae Borealis stars\citep{garcia-hernandez_are_2011}, the peculiar binary XX Oph\citep{evans_solid-phase_2012}, young stellar objects\citep{roberts_detection_2012} and diffuse clouds\citep{berne_detection_2017}. 
Recently, the proposal of C$_{60}$$^+$ as the carrier of two DIBs (9577~\AA~and 9632~\AA) was confirmed by the laboratory spectrum recorded in the gas phase \citep{campbell_laboratory_2015}, which also revealed some weaker absorption features. The weak C$_{60}$$^+$ features were subsequently detected in astronomical spectra \citep{walker_gas-phase_2016,cordiner_confirming_2019}.

Since the first discovery of C$_{60}$ in the gas phase by \cite{kroto_c_1985-1}, the fullerene molecules have been the topic of extensive laboratory studies (see e.g. \cite{bohme_multiply-charged_2011-1,bohme_fullerene_2016,linnartz_c60+_2020-2} and references therein), which provide important knowledge on the possible formation and evolution routes of fullerenes in the Universe.  
For example, the laboratory work showed that C$_{60}$ can be formed starting form the carbon-rich seeded gas via a bottom-up formation route (see e.g. \cite{jager_spectral_2008}). 
On the other hand, the laboratory experiment revealed that C$_{60}$ can be generated from UV radiation induced photochemical evolution of large PAHs \citep{zhen_laboratory_2014}. This may be a possible top-down route formation of C$_{60}$ in the interstellar environment.
The very recent laboratory work suggested that C$_{60}$ can undergo facile formation from shock heating and ion bombardment of circumstellar SiC grains \citep{bernal_formation_2019-1}. 
\cite{garcia-hernandez_infrared_2013-1} summarized and discussed the possible formation mechanisms of fullerene in evolved stars and in ISM in their IR spectroscopic study of C$_{60}$/anthracene adducts.

Fullerenes and their ions are highly active due to the unsaturated features, and thus they can easily react with other molecules \citep{petrie_hydrogenation_1992,murata_reaction_2001,bohme_fullerene_2016,omont_interstellar_2016}. Fullerene/PAHs adducts are one family of the fullerene reaction products and are of astrochemical interest. It is known that fullerene/PAHs adducts are generated via the Diels$-$Alder cycloaddition reactions 
\citep{briggs_[60]fullereneacene_2006,petrie_laboratory_2000,cataldo_sonochemical_2014-1,zhen_laboratory_2019-2}.
On the other hand, anthracene (C$_{14}$H$_{10}$) is among the simple PAH molecules, whose reaction with C$_{60}$ has received considerable attentions. For example, \citet{cataldo_sonochemical_2014-1} reported the sonochemical synthesis of monoanthracene adduct (C$_{14}$H$_{10}$)C$_{60}$ and bis-anthracene adduct (C$_{14}$H$_{10}$)$_2$C$_{60}$ with the precursors of C$_{60}$ and anthracene dissolved in benzene. The same group also studied the IR spectra of these adducts by using Fourier transform IR spectroscopy. More importantly, it is found that the IR spectra of the C$_{60}$/anthracene adducts are similar to those of C$_{60}$ and other unidentified IR emission bands recorded by astronomical observations \citep{garcia-hernandez_infrared_2013-1}. Motivated by \cite{garcia-hernandez_infrared_2013-1}'s work, the formation and photochemistry of (C$_{60}$ and C$_{70}$)/anthracene cluster cations in the gas phase were investigated in this lab \citep{zhen_laboratory_2019-2}. 
The fullerene/anthracene cluster cations are formed from C$_{56/58/60}$$^+$ or C$_{66/68/70}$$^+$ and neutral anthracene molecules via ion-molecule reactions.
Upon irradiation by a 355 nm laser beam, the cluster cations dissociated into fullerene cations and neutral anthracene molecules. Besides, the experimental results showed that C$_{60}$$^+$ and C$_{70}$$^+$ have lower reactivity compared to their neighbor fullerene cations (C$_{56/58}$$^+$ and C$_{68}$$^+$).

It should be mentioned that \citet{candian_searching_2019-1} reported a theoretical study of the stability and IR spectra of neutral and ionized fullerenes with a coverage from C$_{44}$ to C$_{70}$. By comparing the theoretical IR spectra to the observed emission spectra of several planetary nebulae, the authors suggested the possible presence of smaller cages (44, 50 and 56 carbon atoms) in the astronomical objects. 
In this contribution, we present the laboratory study on the formation of cluster cations between fullerene cations (C$_{44}$$^+$ to C$_{70}$$^+$) and anthracene molecules, and investigate 
the reactivity of these fullerenes, especially the ones containing 44, 50 and 56 carbon atoms. In order to generate the interested fullerene cations, higher laser energy and longer irradiation time are used compared to our recent work \citep{zhen_laboratory_2019-2}. The experimental details are given in the following section. The results and discussions, and astronomical implications are presented in section \ref{sec:e results} and section \ref{sec:astronomic implication}, respectively. Conclusions follow in section \ref{sec:conclusions}.

\section{Experimental methods}
\label{sec:exp}
The experiment was performed using the quadrupole ion trap and time-of-flight (QIT-TOF) mass spectrometry setup, which has been described in detail elsewhere \citep{zhen_laboratory_2019-3,zhen_laboratory_2019-2}. Briefly, the gas phase fullerene molecules (C$_{60}$ or C$_{70}$) were prepared by heating their powder samples at a temperature of $\sim$ 613~K and then ionized by electrons ($ \sim $ 82~eV) produced in an electron gun (Jordan, C-950). The interested fullerene cations are selected by using an ion gate and a quadrupole mass filter (Ardara, Quad-925mm-01) and then guided into the quadrupole ion trap (Jordan, C-1251).
Another oven (used at room temperature) mounted under the quadrupole ion trap was used to vaporize anthracene powder. The gas phase anthracene molecules effused continuously towards the center of the ion trap. 
Through the ion-molecule reactions between fullerene cations and anthracene molecules in the ion trap, fullerene/anthracene cluster cations were produced.
The third harmonic output (355 nm) of a Nd:YAG laser (Spectra-Physics, INDI) was used to irradiate the fullerene cations and fullerene/anthracene cluster cations and induce the photochemistry process.
At a appropriate timing, the ions were extracted from the ion trap and detected by a reflection TOF mass spectrometer (Jordan, D-850).

In order to produce smaller fullerenes, we optimized the experimental conditions and found that the laser energy and the irradiation time are critical to the production of smaller fullerene cations. By monitoring the intensity of new generated fragment ions in the mass spectrum, it is found that the optimal conditions are with the laser energy of $ \sim $~30 mJ/pulse and irradiation time of 1.6 s in each measured period.

The simple PAH, anthracene (C$_{14} $H$_{10 }$), was used as the reactant to examine the reactivity of fullerene cations based on the following considerations. The fullerene/anthracene adducts are among the simple and typical fullerene/PAHs adducts. Therefore, the study of formation and photochemical processes of fullerene/anthracene adducts can provide a guidance for other fullerene/PAHs clusters.
Furthermore, the anthracene molecule allows us to make a reasonable comparison with the previous work \citep{zhen_laboratory_2019-2} where the same molecule is used as the reactant.
At last, due to its relatively high vapor pressure at room temperature, the anthracene molecule is suitable for our current experimental setup.

\section{Experimental results and discussions}
\label{sec:e results}

\begin{figure}[t]
	\centering
	\includegraphics[width=\columnwidth]{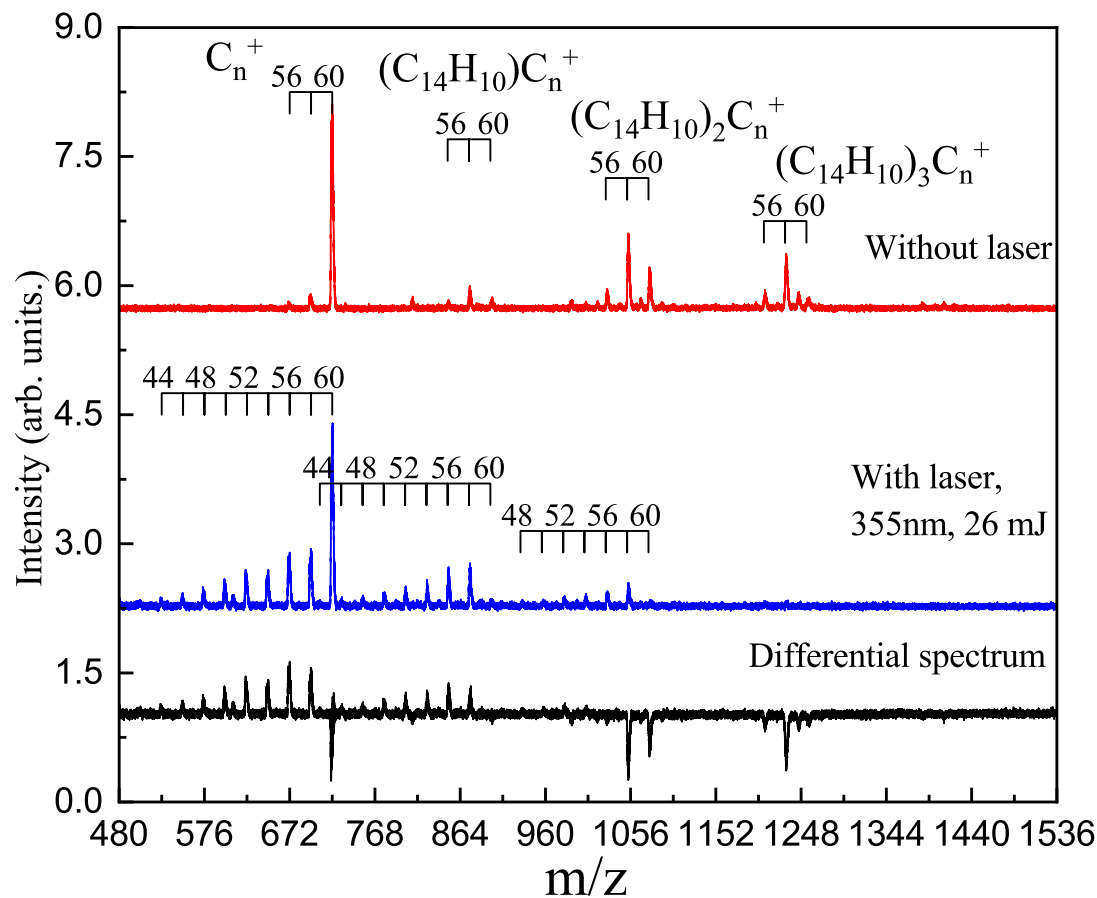}
	\caption{Mass spectrum of fullerene (C$_{60}$)/anthracene cluster cations recorded without laser
		irradiation (top red trace) and with laser irradiation (middle blue trace). We used 355 nm laser with  energy of 26~mJ/pulse and irradiation time amounting to 1.6 s. The  assignments of mass spectral peaks are shown. The differential spectrum between the blue trace (laser on) and the red trace (laser off) is also shown in the bottom trace with black color. 
		}
	\label{fig1}
\end{figure}

The mass spectrum of the fullerene (C$_{60}$)/anthracene cluster cations recorded without laser irradiation is shown in Fig.~\ref{fig1} (top red trace). Fullerene cations (C$_{56/58/60}$$^+$) and series of fullerene/anthracene cluster cations ((C$_{14}$H$_{10}$)$_n$C$_{56/58/60}$$^+$, n=1-3) were produced in the experiment. The fullerene cations were generated by electron bombardment of neutral C$_{60}$ molecules, and the cluster cations were formed via ion-molecule reactions between C$_{56/58/60}$$^+$ and anthracene molecules  in the ion trap. 
Fig.~\ref{fig1} (middle blue trace) shows the recorded mass spectrum of trapped fullerene/anthracene cluster cations after laser irradiation with energy of 26 mJ/pulse and irradiation time of 1.6 s (i.e., typically $ \sim $16 pulses). It can be seen that series of fullerene cations (C$_{(60-2n)}$$^+$, n=0-8) and fullerene/anthracene cluster cations ((C$_{14}$H$_{10}$)C$_{(60-2n)}$$^+$, n=0-8 and (C$_{14}$H$_{10}$)$_2$C$_{(60-2n)}$$^+$, n=0-6) are formed in the ion trap. 
The bottom trace of Fig.~\ref{fig1} displays the differential spectrum to reflect the intensity changes of these cations formed under the conditions of laser-off and laser-on. It can be seen that the intensity of larger mass cluster cations ((C$_{14}$H$_{10}$)$_{2/3}$C$_{56/58/60}$$^+$, (C$_{14}$H$_{10}$)C$_{60}$$^+$) and fullerene cation (C$_{60}$$^+$) decreases after laser irradiation, while the intensity of other cations increases.

\begin{figure*}[t]
	\centering
	\includegraphics[width=\textwidth]{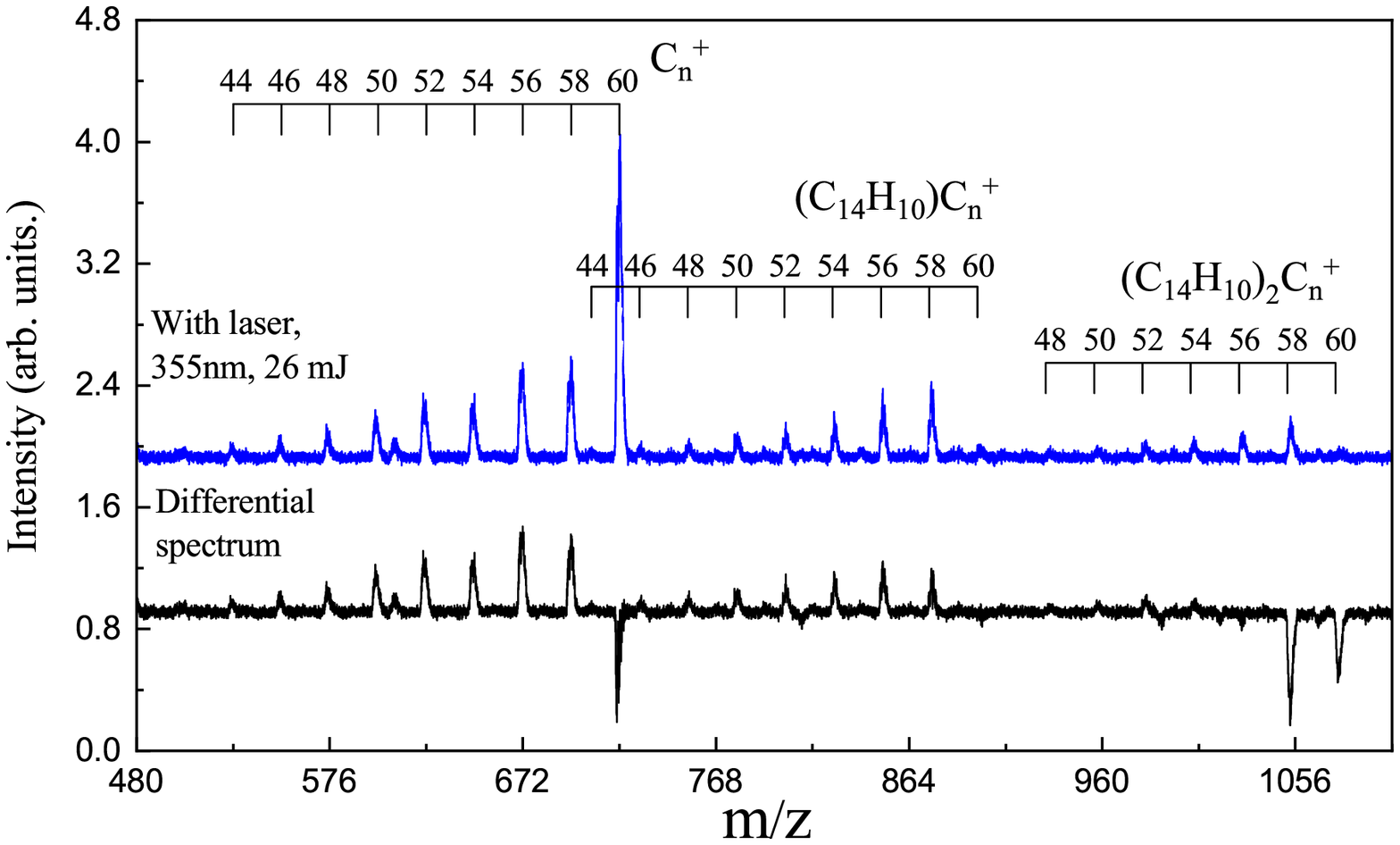}
	\caption{The zoom-in mass spectrum of fullerene (C$_{60}$)/anthracene cluster cations with laser irradiation and the differential spectrum in the range of m/z=480$-$1104. 
	}
	\label{fig2}
\end{figure*}

To clearly show the intensity changes, a zoom-in of the middle and bottom traces of Fig.~\ref{fig1} is displayed in Fig.~\ref{fig2}. 
As the number of carbon atoms decreases, the intensities for fullerene cations, their mono-anthracene adducts and bis-anthracene adducts become weak gradually.  Interestingly, the intensity of (C$_{14}$H$_{10}$)C$_{60}$$^+$ is weaker not only than its neighbor ((C$_{14}$H$_{10}$)C$_{58}$$^+$), but also than other smaller mono-adducts.
Likewise, the intensity of (C$_{14}$H$_{10}$)$_2$C$_{60}$$^+$ is weaker than the other smaller bis-adducts.

In recent study of \cite{zhen_laboratory_2019-2}, laser energy of 1.3~mJ/pulse and irradiation time of 0.5~s were used to irradiate the trapped ions. 
The fullerene cations, C$_{56}$$^+$, C$_{58}$$^+$ and C$_{60}$$^+$, and their corresponding fullerene/anthracene cluster cations were recorded in the mass spectra (Fig.~2 in Ref. \cite{zhen_laboratory_2019-2}).
By contrast, more smaller fullerene cations (C$_{(60-2n)}$$^+$, n=3-8) and fullerene/anthracene cluster cations ((C$_{14}$H$_{10}$)C$_{(60-2n)}$$^+$, n=3-8 and (C$_{14}$H$_{10}$)$_2$C$_{(60-2n)}$$^+$, n=3-6) were newly observed in current mass spectra (Fig.~\ref{fig1}).
Note that the laser wavelength (355~nm) used in these two studies was the same. The newly observed smaller fullerene cations and their fullerene/anthracene cluster cations were due to the higher energy (26~mJ/pulse) and longer irradiation time (1.6~s) used in present experiment.

The observed smaller fullerene cations in present study (Fig.~\ref{fig1}) all contain even numbers of carbon atoms, indicating the successive C$ _2 $ loss in the formation process, which has been known as the main evolution way followed by fullerene cations after absorption of photons \citep{lifshitz_c2_2000,zhen_laboratory_2014}. Combining this consideration with previous studies \citep{zhen_laboratory_2014, zhen_laboratory_2019-4,zhen_laboratory_2019-2}, the following photochemistry mechanism (path~1-4) is suggested.
After absorption of UV photons, fullerene/anthracene cluster cations dissociated into fullerene cations and anthracene molecules (path~\ref{equ:C60_1}). And then, the fullerene cations including free ones and ones resulted from the dissociation underwent successive C$_2$ loss and formed smaller fullerene cations (path~\ref{equ:C60_2}). After that, the smaller fullerene cations reacted with anthracene molecules to form fullerene/anthracene cluster cations (path~\ref{equ:C60_3} and \ref{equ:C60_4}). Reaction pathways (in sequence) for the formed fullerene/anthracene cluster cations are summarized as below:

\begin{small}
\begin{eqnarray}
\label{equ:C60_1}	&[\rm {(C_{14}H_{10}){_{(1-3)}}C_{56/58/60}}]^+ \stackrel{\rm h\nu}{\longrightarrow} \rm (C_{14}H_{10}){_{(1-3)}} + \rm [C_{56/58/60}]^+\\
\label{equ:C60_2}	&[\rm C_{56/58/60}]^+ \stackrel{\rm h\nu}{\longrightarrow} \rm nC_{2} + \rm [C_{(60-2n)}]^+\\
\label{equ:C60_3}	&[\rm C_{(60-2n)}]^+ + \rm C_{14}H_{10}{\longrightarrow} [\rm (C_{14}H_{10})C_{(60-2n)}]^+\\
\label{equ:C60_4}	&[\rm (C_{14}H_{10})C_{(60-2n)}]^+ + \rm C_{14}H_{10}\longrightarrow \rm [(C_{14}H_{10}){_2}C_{(60-2n)}]^+
\end{eqnarray}
\end{small}

\begin{figure}[t]
	\centering
	\includegraphics[width=\columnwidth]{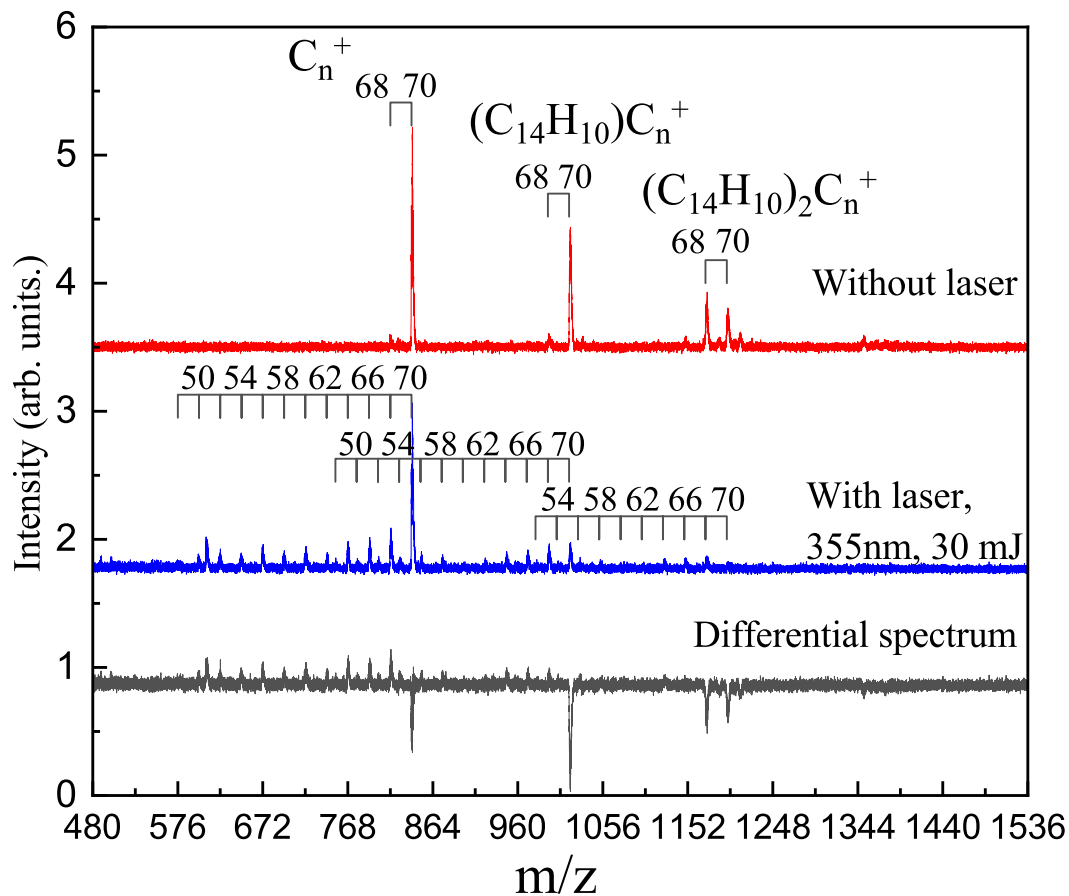}
	\caption{Mass spectrum of fullerene (C$_{70}$)/anthracene cluster cations recorded without laser irradiation (top red trace) and with laser irradiation (middle blue trace). We used 355 nm laser with energy  of 30~mJ/pulse and irradiation time amounting to 1.6 s. The  assignments of mass spectral peaks are shown. The differential spectrum between the blue trace (laser on) and the red trace (laser off) is also shown in the bottom trace with black color. 
	}
	\label{fig3}
\end{figure}

In addition to C$ _{60} $, C$_{70}$ was also used as the precursor to produce the smaller fullerene cations and examine their reactivity with anthracene. The experiment conditions were same as those used for the study of C$ _{60} $ except for the laser energy of 30 mJ/pulse. The recorded mass spectra are depicted in Fig.~\ref{fig3}. 
The top red trace shows the mass spectrum of the C$_{70}$/anthracene cluster cations recorded without laser irradiation. Fullerene cations (C$_{68/70}$$^+$) and their anthracene cluster cations ((C$_{14}$H$_{10}$)$_m$C$_{68/70}$$^+$, m=1,2 were observed. 
After irradiation of these trapped ions by the 355 nm laser beam with energy of 30 mJ/pulse and irradiation time of 1.6 s, the recorded mass spectrum is displayed as the middle blue trace in Fig.~\ref{fig3}.
Series of fullerene cations (C$_{(70-2m)}$$^+$, m=2-11), their mono-anthracene adducts ((C$_{14}$H$_{10}$)C$_{(70-2m)}$$^+$, m=2-11 and bis-anthracene adducts (C$_{14}$H$_{10}$)$_2$C$_{(70-2m)}$$^+$, m=2-9) were newly formed comparing to the top trace without laser irradiation. 
In order to clearly show the changes in the mass spectra recorded without and with laser irradiation, the differential spectrum is derived by extraction the top trace (without laser beam) from the middle trace (with laser beam). The resultant spectrum is shown as the bottom trace in Fig.~\ref{fig3}. The peaks with positive intensity indicate that the cations were formed after laser irradiation, such as C$_{(70-2m)}$$^+$, m=2-11.

\begin{figure*}[t]
	\centering
	\includegraphics[width=\textwidth]{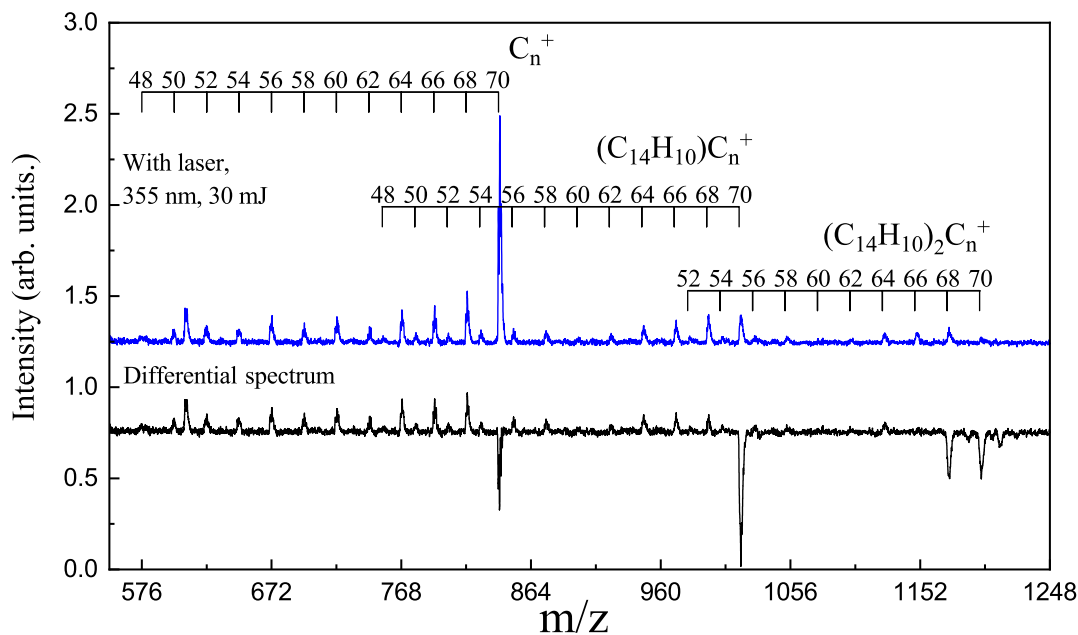}
	\caption{The zoom-in mass spectrum of fullerene (C$_{70}$)/anthracene cluster cations with laser irradiation and the differential spectrum in the range of m/z=552$-$1248. 
	}
	\label{fig4}
\end{figure*}

Fig.~\ref{fig4} displays the zoom-in of the middle and bottom traces of Fig.~\ref{fig3}. It can be seen that the intensities for fullerene cations, their mono-anthracene adducts and bis-anthracene adducts became weak gradually as the carbon numbers get smaller. The intensity of (C$_{14}$H$_{10}$)C$_{60}$$^+$ is weaker than its neighbors, (C$_{14}$H$_{10}$)C$_{58}$$^+$ and (C$_{14}$H$_{10}$)C$_{62}$$^+$. 
Likewise, the intensity of (C$_{14}$H$_{10}$)$_2$C$_{60}$$^+$ is weaker than its neighbors, (C$_{14}$H$_{10}$)$_2$C$_{58}$$^+$ and (C$_{14}$H$_{10}$)$_2$C$_{62}$$^+$.

The increments in laser energy and irradiation time on the C$_{70}$$^+$/anthracene system resulted in the production of more smaller fullerene/anthracene cluster cations (see Fig.~\ref{fig3}), which is similar as the C$_{60}$$^+$/anthracene system (Fig.~\ref{fig1}).
In the previous work \citep{zhen_laboratory_2019-2}, laser energy of 0.9~mJ/pulse and irradiation time of 0.5~s were used to irradiate the trapped C$_{70}$$^+$/anthracene cluster cations. The fullerene cations, C$_{70}$$^+$ and C$_{68}$$^+$, and their corresponding fullerene/anthracene cluster cations were recorded in the mass spectra (Fig.~3 in Ref. \cite{zhen_laboratory_2019-2}). 
Compared to those results, more smaller fullerene cations (C$_{(70-2m)}$$^+$, m=2-11), their mono-anthracene adducts ((C$_{14}$H$_{10}$)C$_{(70-2m)}$$^+$, m=2-11) and bis-anthracene adducts ((C$_{14}$H$_{10}$)$_2$C$_{(70-2m)}$$^+$, m=2-9) were newly observed in current mass spectra (Fig.~\ref{fig3}).
Note that the laser wavelength (355~nm) used in these two studies is same. The newly observed smaller fullerene cations and their fullerene/anthracene cluster cations can be attributed to the higher energy (30~mJ/pulse) and longer irradiation time (1.6~s ) used in present experiment. 

Irradiation UV photons upon the C$_{70}$$^+$/anthracene system, the generated smaller fullerene cations and their anthracene adducts all contain even numbers of carbon atoms. It is suggested that the successive C$ _2 $ loss is dominated the evolution process.
After absorption of UV photons, fullerene/anthracene cluster cations ((C$_{14}$H$_{10}$)$_m$C$_{68/70}$$^+$, m=1,2) dissociated into fullerene cations and anthracene (path~\ref{equ:C70_1}). 
Then, the fullerene cations (including the free ones) underwent photo-fragmentation process (successive C$_2$ loss) and produced serials of smaller fullerene cations (path~\ref{equ:C70_2}). 
Subsequently, these smaller fullerene cations reacted with anthracene molecules to form fullerene/anthracene cluster cations again (path~\ref{equ:C70_3} and \ref{equ:C70_4}). 
The photochemical pathways for the C$_{68/70}$/anthracene cluster cations are summarized as below:
\begin{eqnarray}
\label{equ:C70_1}	& \rm [({C_{14}H_{10}){_{(1-2)}}C_{68/70}}]^+ \stackrel{ \rm h\nu}{\longrightarrow} \rm (C_{14}H_{10}){_{(1-2)}} + \rm [C_{68/70}]^+\\ 
\label{equ:C70_2}	&\rm [C_{68/70}]^+ \stackrel{\rm h\nu}{\longrightarrow}  \rm mC_{2} + \rm [C_{(70-2m)}]^+\\ 
\label{equ:C70_3}	&\rm [C_{(70-2m)}]^+ + \rm {C_{14}H_{10}}{\longrightarrow} \rm [(C_{14}H_{10})C_{(70-2m)}]^+\\ 
\label{equ:C70_4}	&\rm [(C_{14}H_{10})C_{(70-2m)}]^+ + \rm {C_{14}H_{10}}{\longrightarrow} \rm [(C_{14}H_{10}){_2}C_{(70-2m)}]^+ 
\end{eqnarray}

\begin{figure*}[t]
	\centering
	\includegraphics[width=\textwidth]{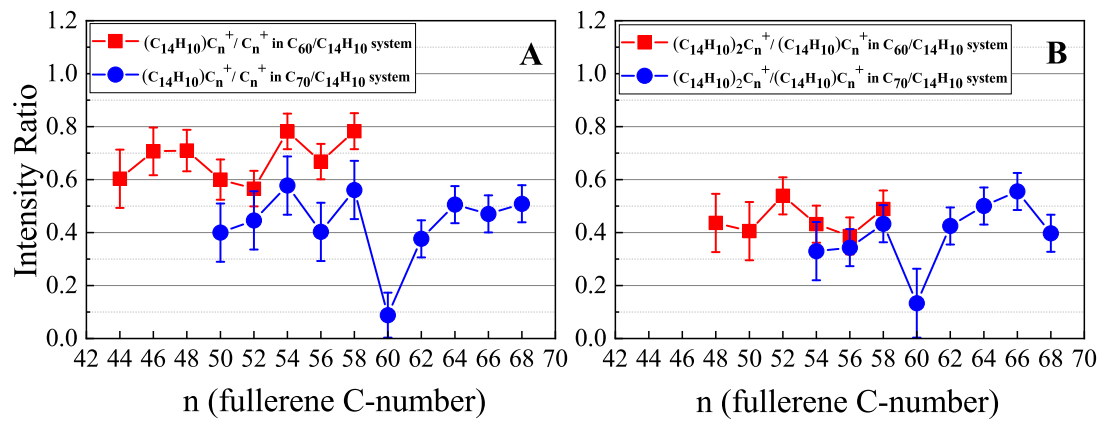}
	\caption{Panel (A):the intensity ratio of formed (anthracene)fullerene cluster cations to fullerene cations in the irradiated spectrum: red line is for the C$_{60}$/anthracene system; blue line is for the C$_{70}$/anthracene system; Panel (B):the intensity ratio of formed (anthracene)$_2$fullerene cluster cations to (anthracene)fullerene cations in the irradiated spectrum: red line is for the C$_{60}$/anthracene system; blue line is for the C$_{70}$/anthracene system.
	}
	\label{fig5}
\end{figure*}

The relative reactivity of fullerene cations can be derived from the intensity variations of fullerene cations and their corresponding fullerene/anthracene cluster cations in the recorded mass spectra.
To illustrate this, Fig.~\ref{fig5}(A) depicts the intensity ratios for (anthracene)fullerene cluster cations to fullerene cations recorded in the C$_{60}$/anthracene system and C$_{70}$/anthracene system individually.
Fig.~\ref{fig5}(B) displays the intensity ratios for (anthracene)$_2$fullerene cluster cations to (anthracene)fullerene cluster cations recorded in the C$_{60}$/anthracene system and C$_{70}$/anthracene system individually.
The peak intensity in the recorded mass spectra (Fig.~\ref{fig2} and \ref{fig4}) were used in the calculation.
The intensity ratio calculated in this way is a reflection of the proportion that fullerene cations or (anthracene)fullerene cluster cations have reacted with free anthracene molecules. In other words, the larger the intensity ratio is, the higher reactivity the corresponding fullerene cation has. 
As shown in Fig.~\ref{fig5}(A), the intensity ratios for C$_{60}$/anthracene system are in a similar range with a central value of $\approx$0.7.  The intensity ratios of C$_{70}$/anthracene system are in a similar range with a central value of $\approx$0.5 with the exception of C$_{60}$$^+$ whose value is about 0.1. 
Similar trends are found in Fig.~\ref{fig5}(B), the intensity ratio for (C$_{14}$H$_{10}$)C$_{60}$$^+$ is of $\approx$ 0.1, which value is obvious smaller than the others. 
The significant lower values of the intensity ratios for C$_{60}$$^+$ and (C$_{14}$H$_{10}$)C$_{60}$$^+$ illustrates that they have lower reactivity in addition reaction with anthracene molecules compared to the other fullerene cations. 
These results confirm the conclusion that C$_{60}$$^+$ has a lower reactivity reported in Ref. \cite{zhen_laboratory_2019-2} where the comparison is limited to C$_{56}$$^+$, C$_{58}$$^+$ and C$_{60}$$^+$ ions.

Despite it is not as apparent as that of C$_{60}$$^+$ cation, there exists a slight variation of intensity ratio for other fullerene cations. As shown in Fig.~\ref{fig5}(A), the intensity ratio for C$_{52}$$^+$ appears as a local minimum in both C$_{60}$/anthracene and C$_{70}$/anthracene systems. Similar thing happens to C$_{44}$$^+$, C$_{50}$$^+$ and probably C$_{56}$$^+$. 
Considering that the mass spectra (Fig.~\ref{fig2} and \ref{fig4}) were recorded under an optimized and stable condition and that 200 average times were taken for each spectrum, the experimental factors (such as laser energy) that caused the intensity ratio variation should have been avoided. Moreover, the variation trends of intensity ratios are observed to be consistent on both C$_{60}$/anthracene and C$_{70}$/anthracene systems (Fig.~\ref{fig5}(A)). 
If the slight variation is not caused by the experiment system errors, it indicates that, within the smaller fullerene cations (44-58 carbon atoms), C$_{44}$$^+$, C$_{50}$$^+$, C$_{52}$$^+$ and probably C$_{56}$$^+$ have relatively lower reactivity towards anthracene molecules. 
It should be mentioned that previous studies \citep{manolopoulos_theoretical_1991-1,zimmerman_magic_1991,rohlfing_production_1984} have shown that the C$_{44} $, C$_{50} $ and C$_{56} $ are more stable than their neighbors. The recent theoretical work \citep{candian_searching_2019-1} addressed the stability of different isomers of C$_{44}$$^+$, C$_{50}$$^+$ and C$_{56}$$^+$. In their study, the stability is characterized in terms of the standard enthalpy of formation per CC bond, the HOMO$-$LUMO gap, and the energy required to eliminate a C$_2 $ fragment. 
These studies \citep{manolopoulos_theoretical_1991-1,zimmerman_magic_1991,rohlfing_production_1984, candian_searching_2019-1} may explain the slight variation observed in present work.

\section{Astronomical implications}
\label{sec:astronomic implication}

The formation and evolution of neutral and charge state fullerenes in astronomical environments are of considerable interest, especially after the detection of C$_{60}$ and C$_{70}$ in the young planetary nebula Tc 1 via their IR emissions \citep{cami_detection_2010-1}. To the best of our knowledge, it is still an open issue so far. 
In this work, we present the laboratory study of the formation and relative reactivity of fullerenes cations (C$_{44}$$^+$ to C$_{70}$$^+$). It is found that the higher laser energy and longer irradiation time are key factors to produce smaller fullerene cations. 
The increments of laser energy and irradiation time essentially increase photon number density.
After absorption of UV photons, the fullerene cations undergo successive C$ _\text{2} $ fragment loss dissociation which results in formation of smaller fullerene cations. 
It indicates that under the strong UV radiation environment like photon-dominated regions, the fullerene cations are probably driven to dissociate to smaller ones or even damage totally. That is to say the top-down evolution of fullerene cations is expected to be dominated in such region. 
However, if the fullerene molecules are on the surface of dust grains, they may survive from the strong irradiation. The case in nebula Tc~1 may be a good example. 
By considering the temperature difference between the observed IR spectra and gas-phase environment in the nebula Tc~1, \citet{cami_detection_2010-1} concluded that C$_{60}$ and C$_{70}$ are in direct contact with solid materials. The solid materials probably shield much UV radiation for the attached fullerenes. As a contrast, the gaseous fullerenes are exposed in the strong radiation environment, and are probably driven to smaller ones or damage. This may be a possible explanation for that no gaseous C$_{60}$ and C$_{70}$ are observed in Tc~1.

The second question addressed by this work is the relative reactivity of fullerene cations (C$_{(60-2n)}$$^+$, n=0-8). The anthracene molecule is used as the reactant. It is found that the smaller fullerene cations show significantly high reactivity when compared to C$_{60}$$^+$.
In the harsh ISM environment, anthracene molecules are not expected to survive. The most abundant interstellar PAHs are large condense ones \citep{ricca_infrared_2012} which are highly reactive as anthracene.  
As is well known, the Diels--Alder adduction is one of the routine ways when fullerene react with PAHs. 
In view of this point, other PAHs are expected to react with smaller fullerene cations in an efficient way like anthracene molecules.
If the smaller fullerene and PAHs co-exist in the same astronomical region, they may react and form fullerene/PAHs adducts. 
These adducts may accumulate and contribute somewhat to the unidentified IR emission features. It is promising when noting the fact that  C$_{60}$/anthracene adducts were shown to have strikingly similar spectral features to those from C$_{60}$ (and C$_{70}$) fullerenes and other unidentified infrared emission features \citep{garcia-hernandez_infrared_2013-1}.

\section{Conclusion}
\label{sec:conclusions}
The formation and relative reactivity of fullerene cations (from smaller to larger, C$_{44}$$^+$ to C$_{70}$$^+$) are studied in this work. It is found that the higher laser energy and longer irradiation time are key factors to produce the smaller fullerene cations such as C$_{(60-2n)}$$^+$, n = 1-8. 
When reacting with anthracene molecules, slightly different reactivity may exist within the smaller fullerene cations, if the experimental system errors are ruled out. Nevertheless, the smaller fullerene cations have significantly higher reactivity compared to C$_{60}$$^+$. 
A successive loss of C$_2 $ fragments mechanism is suggested to account for the formation of smaller fullerene cations, which then undergo addition reaction with anthracene molecules. 
These results obtained here provide a growth route towards fullerene (from smaller to larger in size) derivatives based on PAH-related molecules in a bottom-up growth process and an insight for their photo-evolution behavior in the ISM. The results also suggest, when conditions are favorable, fullerene derivatives can form efficiently.

\section*{Acknowledgements}

This work is supported by the Fundamental Research Funds for the Central Universities, the National Science Foundation of China (NSFC, Grant No. 11743004). We thank the referee for the very constructive and detailed comments which help improve this work a lot. 

\bibliographystyle{raa}
\bibliography{ms-RAA-2020-0011-reference}
\label{lastpage}

\end{document}